\documentclass[a4paper, aps, prb, twocolumn, showpacs, superscriptaddress]{revtex4-1}

\usepackage{amssymb}
\usepackage{amsmath}
\usepackage{color}
\usepackage{graphics, graphicx}
\usepackage{psfrag}
\usepackage{mathcomp}

\newcommand{\IM}{\mbox{Im}}

\begin{document}

\author{D. Semmler} 
\affiliation{Institut f\"ur Theoretische Physik, Johann Wolfgang Goethe-Universit\"at, 60438 Frankfurt am Main, Germany}
\author{K. Byczuk}
\affiliation{Institute of Theoretical Physics, Faculty of Physics, University of Warsaw, ul. Ho\.{z}a 69, 00-681 Warszawa, Poland}
\author{W. Hofstetter}
\affiliation{Institut f\"ur Theoretische Physik, Johann Wolfgang Goethe-Universit\"at, 60438 Frankfurt am Main, Germany}

\date{\today}

\title{Anderson-Hubbard model with box disorder: Statistical dynamical mean-field theory investigation}

\begin{abstract}
Strongly correlated electrons with box disorder in high-dimensional lattices are investigated. We apply the statistical dynamical mean-field theory, which treats local correlations non-perturbatively. The incorporation of a finite lattice connectivity allows for the detection of disorder-induced localization via the probability distribution function of the local density of states. We obtain a complete paramagnetic ground state phase diagram and find correlation-induced as well as disorder-induced metal-insulator transitions. Our results qualitatively confirm predictions obtained by typical medium theory. Moreover, we find that the probability distribution function of the local density of states in the metallic phase strongly deviates from a log-normal distribution as found for the non-interacting case.
\end{abstract}

\maketitle

\section{Introduction}

Anderson localization\cite{Anderson1958,Kramer1993} as well as strong correlation effects due to interactions\cite{Mott1990,Imada1998} have been studied intensely for decades in solid state physics. It is well established that both phenomena modify the transport properties of solids in a crucial way. In particular, disorder as well as correlations are capable of inducing metal-insulator transitions known as the Anderson transition and the Mott-Hubbard transition, respectively. 
Although each fundamental phenomenon - correlations and disorder - on its own gives rise to challenges to both experimental and theoretical research and is still a subject of current investigations, the realistic modeling of materials requires the simultaneous consideration of both effects. This interplay is of central interest within present day's solid state physics.\cite{Lee1985,Altshuler1985,Belitz1994,Kravchenko2004,Miranda2005} 

Theoretical investigations, e.g. of the Anderson-Hubbard model combining the Hubbard model and the Anderson model are notoriously difficult. The most intriguing effects, such as the Mott transition and the Anderson transition, take place at intermediate interaction and disorder strength requiring a non-perturbative approach. Extensions of the dynamical mean-field theory (DMFT)\cite{Metzner1989,Georges1996,Kotliar2006} to disordered systems are especially promising in this respect, since local correlations are incorporated non-perturbatively.
Such extensions were performed on a level of the coherent potential approximation\cite{Vlaming1992,Ulmke1995} and on a level of typical medium theory (TMT) incorporating localization effects.\cite{Dobrosavljevic2003,Byczuk2005,Aguiar2006,Aguiar2009,Byczuk2009,Byczuk2010,Dobrosavljevic2010} A shortcoming of both approaches is given by the non-consideration of disorder-induced fluctuations of local observables. Dobrosavljevi\'{c} and Kotliar extended the DMFT to include disorder fluctuations in space on a fully stochastic level,\cite{Dobrosavljevic1997} which is referred to as statistical DMFT. \cite{Miranda2001,Aguiar2003,Bronold2004,Miranda2005,Tran2007,Song2008,Song2009,Semmler2010,Semmler2010b} This theory is capable of describing Mott and Anderson-Mott transitions non-perturbatively and allows for the self-consistent determination of the probability distribution function (PDF) of local observables. 

Among the investigations of the box disordered Anderson-Hubbard model in $d>1$ dimensions which are non-perturbative and incorporate localization effects the TMT was used in the most comprehensive way.\cite{Byczuk2005,Aguiar2009,Byczuk2009,Byczuk2010,Dobrosavljevic2010} Within these works the ground state phase diagram has been determined for a half-filled band \cite{Byczuk2005,Aguiar2009} and the physical two fluid picture of the Anderson-Mott transition has been developed.\cite{Aguiar2009,Dobrosavljevic2010} The statistical DMFT has also been used to study the box disordered Anderson-Hubbard model.\cite{Dobrosavljevic1997,Song2008,Song2009} The Anderson-Mott transition has been characterized in detail by an accurate treatment of the low energy excitations.\cite{Dobrosavljevic1997} Remarkably, two critical disorder strengths have been found, one corresponding to the entering of a non-Fermi liquid phase and a second one corresponding to the Anderson-Mott transition. A special focus was laid on the zero-bias anomaly \cite{Efros1975,Altshuler1979,Chiesa2008} in two dimensions.\cite{Song2008,Song2009} However, partially due to the immense numerical effort and partially due to the restrictions of the used impurity solvers -- the slave-boson approach\cite{Kotliar1986,Georges1996} and the Hubbard-I approximation\cite{Hubbard1963} -- the ground state phase diagram was not determined. 

In this work we revisit the Anderson-Hubbard model with box disorder by means of the statistical DMFT. By using the modified perturbation theory (MPT)\cite{Kajueter1996,Potthof1997} as an impurity solver we determine the PDF of the local density of states (LDOS) for finite interaction strengths which was not done before. Our MPT investigation shows that the PDF of the LDOS is not well fitted by a log-normal distribution function within the metallic phase in contrast to the non-interacting case.\cite{Mirlin1996,Mirlin2000,Schubert2010} Finally, the paramagnetic ground state phase diagram consisting of disordered correlated metal, Mott insulator, and Anderson-Mott insulator is determined and critically discussed with regard to established results obtained by means of TMT earlier.\cite{Byczuk2005,Aguiar2009,Byczuk2010,Dobrosavljevic2010}

\section{Model}\label{system}
A tight-binding model that describes  strong correlations and randomness is given by the Anderson-Hubbard Hamiltonian
\begin{equation}
 H = -\sum\limits _{ij\sigma} t_{ij} c_{i \sigma}^{\dagger} c_{j \sigma}   
 - \sum \limits_{i \sigma} (\mu - \epsilon_i) c_{i \sigma}^{\dagger} c_{i \sigma} + U \sum \limits_{i}  n_{i \uparrow } n_{i \downarrow } \;, \label{eq_sys1}
\end{equation}
where $c_{i \sigma}^{\dagger}$  ($c_{i \sigma}$) denotes the fermionic creation (annihilation) operator at a lattice site $i$ with spin $\sigma=\pm 1/2$. The number operator is given by $n_{i \sigma}= c_{i \sigma}^{\dagger}c_{i \sigma}$. The hopping amplitude $t_{ij}=t$ for nearest neighbor (n.n.) sites $i$ and $j$ only and zero otherwise, the local interaction strength is denoted by $U$, and $\mu$ represents the chemical potential. In the following, we consider electrons on the Bethe lattice),\cite{Georges1996,Eckstein2005,Kollar2005} which is characterized by the connectivity $K$. It is related to the lattice coordination number $z$ via $K=z-1$. 

Within the tight-binding model the disorder manifests itself by random on-site energies $\epsilon_i$, which are drawn from a PDF $p_{\epsilon}(\epsilon _i)$. Here, we consider the box distribution 
\begin{equation}
p_{\epsilon}(\epsilon _i) = \frac {1}{\Delta} \Theta \big(\frac{ \Delta}{2} - | \epsilon _i | \big)  \;, \label{PD_onsite_E}
\end{equation}
which is commonly used model in theoretical investigations. Therein, the disorder strength is denoted by $\Delta$ and we assume that the on-site energies of all lattice sites are independently and identically distributed.

\section{Statistical DMFT}\label{method}

The statistical DMFT\cite{Dobrosavljevic1997} is an extension of the well-known DMFT to cope with disordered systems, which enables the self-consistent determination of the PDF of the local single-particle Green's functions $p\left[G_{ii\sigma}(\omega)\right]$. Here, $G_{ij\sigma}(\omega)$ is the Fourier transform of the retarded Green's function $G_{ij\sigma}(t)=-i\Theta(t)\langle [ c_{i\sigma}(t),c_{j\sigma}^{\dagger}(0) ]_+ \rangle$, where $[..,..]_+$ represents the anticommutator bracket. In the discussion that follows, the spin index $\sigma$ is omitted, since we consider paramagnetic solutions of the Anderson-Hubbard model. 

In the non-interacting system, the renormalized perturbation theory\cite{Economou_book} shows that the local Green's function is given by
\begin{equation}
G_{ii} (\omega) = \frac{1}{\omega -\epsilon_i - \Gamma_i(\omega) + i \eta} \,, \label{eq_M1}
\end{equation} 
where the hybridization function $\Gamma_i(\omega)$ describes the coupling of site $i$ to the nearest neighbor lattice sites. The introduction of a finite broadening factor $\eta>0$ is essential for the numerical distinction between localized and extended states as discussed in detail later. The hybridization function can be expressed by an infinite, renormalized series 
\begin{eqnarray}
\Gamma_i (\omega) &=&  \sum\limits_{j\neq i} t_{ij} G_{jj}^{(i)}(\omega)t_{ji} \nonumber\\ &&
+ \sum\limits_{j\neq i; k\neq j,i} t_{ij} G_{jj}^{(i)}(\omega)t_{jk} G_{kk}^{(j,i)}(\omega)t_{ki}+ \ldots ,\label{eq_LD2}
\end{eqnarray} 
where $G_{qq}^{(n,m,\ldots)}(\omega)$ is the diagonal cavity Green's function of the system when the sites $n,m,\ldots$ are removed. The corresponding diagonal cavity Green's functions are determined by similar series on cavity lattices leading to a hierarchy of equations.

On the Bethe lattice with connectivity $K$ this series is exactly truncated after the first term and the hybridization functions are exactly given by\cite{Abou-Chacra1973,Eckstein2005,Kollar2005,Economou_book}
\begin{eqnarray}
\Gamma_i (\omega) &=&  \sum\limits_{j\;\mbox{\footnotesize n.n. of}\;i} t_{ij} \frac{1}{\omega-\epsilon_j-\Gamma_j^{(i)}(\omega)+i\eta} t_{ji}  \\
\Gamma_j^{(i)} (\omega) &=&  \sum\limits_{k\;\mbox{\footnotesize n.n. of}\;j} t_{jk} \frac{1}{\omega-\epsilon_k-\Gamma_k^{(i,j)}(\omega)+i\eta} t_{kj}  \\
\Gamma_k^{(i,j)} (\omega) &=& \sum\limits_{l\;\mbox{\footnotesize n.n. of}\;k} t_{kl} \frac{1}{\omega-\epsilon_l-\Gamma_l^{(i,j,k)}(\omega)+i\eta} t_{lk} \\
\ldots \nonumber
\end{eqnarray}
Additionally, the cavity hybridization functions of lattices with several sites removed are also simplified due to the absence of loops on the Bethe lattice. The corresponding equations reproduce their structure from the second equation and therefore there are only two classes of hybridization functions (and corresponding Green's functions) determined by the equations
\begin{eqnarray}
\Gamma_i(\omega) &=& \sum\limits_{j=1}^z t_{ij}^2 G_{jj}^{(i)}(\omega) \label{ld_Bethe_hyb_eq1}\\  
\Gamma_j^{(i)}(\omega) &=& \sum\limits_{k=1}^K t_{jk}^2 G_{kk}^{(j)}(\omega) \;.  \label{ld_Bethe_hyb_eq2}
\end{eqnarray} 
Here, both sums extend over the nearest neighbors and the second one is defined on the cavity lattice. Next, we employ  the further approximation that was introduced and successfully used by Abou-Chacra \textit{et al.}\cite{Abou-Chacra1973,Abou-Chacra1974} Therein, the structural differences in the equations (\ref{ld_Bethe_hyb_eq1}) and (\ref{ld_Bethe_hyb_eq2}) are neglected and replaced by a single equation
\begin{equation}
\Gamma_i(\omega) =  \sum\limits_{j=1}^K t_{ij}^2 G_{jj}(\omega) \;. \label{eq_hybrid}
\end{equation}
This approximation is better the higher the connectivity is, since the difference between $z$ and $K$ becomes smaller. Furthermore, we note that all equations in the hierarchy incorporate the sum over $K$ diagonal Green's Functions except the first.\cite{comment1} Thus, $K$ can be considered as a typical number of summands in the equations of the hierarchy.\cite{Abou-Chacra1973} 

Following Dobrosavljevi\'{c} and Kotliar\cite{Dobrosavljevic1997} for finite interactions, an additional local self-energy $\Sigma_{i}(\omega)$, taking into account the interaction effects, is incorporated via $G_{ij} = G_{ij}^0 [\epsilon_i\rightarrow \epsilon_i +\Sigma_i]$, where $G_{ij}^0$ corresponds to the Green's function for the same disorder realization in the non-interacting case. The local single-particle Green's function is consequently given by 
\begin{equation}
G_{ii} (\omega) = \frac{1}{\omega + \mu -\epsilon_i - \Sigma_i(\omega) - \Gamma_i(\omega) + i \eta} \,. \label{eq_M4}
\end{equation} 
The approximation of a local self-energy is exact for infinite connectivity, as was shown by Metzner and Vollhardt in a seminal paper\cite{Metzner1989} and is known to be a good approximation in three spatial dimensions. Since the translational invariance is broken in a disordered system, the local single-particle Green's function becomes site-dependent, giving rise to a PDF $p\left[G_{ii}(\omega)\right]$. Within statistical DMFT, this PDF is determined by an ensemble with a large number $N$ of Green's functions. Each of these Green's functions corresponds to an impurity problem defined by a site-dependent hybridization function $\Gamma_{i}(\omega)$. The solution of each impurity problem results in local self-energies $\Sigma_i(\omega)$.\cite{Georges1996,Dobrosavljevic1997} Hence, the statistical DMFT maps the original lattice model onto an ensemble of impurities, whose PDF has to be determined self-consistently. The statistical DMFT naturally adopts Anderson's point of view\cite{Anderson1958} that an investigation of disordered systems should focus on the PDFs of local observables.

The self-consistent calculation scheme is implemented by the following algorithm: The starting point is an initial PDF $p\left[G_{ii}(\omega)\right]$ and the calculation is performed by (i)~drawing a random on-site energy $\epsilon_i$ from the PDF $p_{\epsilon}(\epsilon_i)$ given in Eq.~(\ref{PD_onsite_E}) for each sample of the ensemble; (ii) The hybridization function $\Gamma_i(\omega)$, with the local single-particle Green's function $G_{jj}(\omega)$ of the nearest neighbors randomly sampled from the PDF $p\left[G_{ii}(\omega)\right]$, is determined via Eq.~(\ref{eq_hybrid}) for each sample; (iii) the local self-energy $\Sigma_i(\omega)$ is calculated from the solution of the local impurity problem by using an impurity solver; (iv) the local single-particle Green's function $G_{ii}(\omega)$ is calculated using Eq.~(\ref{eq_M4}); (v) having calculated a completely new ensemble $\{ G_{ii}(\omega)\}$, a new PDF $p\left[G_{ii}(\omega)\right]$ is obtained and we return to step (i). The steps (i)-(v) are to be repeated until self-consistency for the PDF $p\left[ G_{ii}(\omega)\right]$ is achieved. We note that this method incorporates spatial fluctuations, i.e. quantum interference effects via  (ii). Statistical fluctuations can be minimized by an artificial increase of the size of the ensemble after having reached self-consistency. In this work, typically $80-100$ successive DMFT iterations are used.

One physical observable that is capable of describing the Mott transition and the Anderson-Mott transition on equal footing is given by the LDOS.\cite{Dobrosavljevic2010,Semmler2010} It is defined by $\rho_i (\omega) = - \frac{1}{\pi} \IM(G_{ii}(\omega))$ and is a random quantity in disordered systems. The corresponding PDF $p[\rho_i (\omega)]$ is obtained by counting all values of the LDOS for each frequency and by constructing a histogram.  
From the calculated PDF we can extract all moments of the PDF of the LDOS  
\begin{equation}
M^{(k)}_{\rho(\omega)} := \int\limits_0^{\infty} d\rho'\; p[\rho'(\omega)] \rho'^k \label{eq_p_moments}.
\end{equation}
In particular, the first moment ($k=1$) is the arithmetic average of the LDOS
\begin{equation}
\langle \rho (\omega) \rangle_{\mbox{\tiny arith}} = \int  d\rho'(\omega) \; p[\rho'(\omega)] \, \rho'(\omega) \label{eq_arith_av}\;,
\end{equation}
where the dependence on $\omega$ is to be understood parametrically. The arithmetic average of the LDOS corresponds to the density of states  (DOS) of the system. Furthermore, the typical value, i.e. the most probable value, can be extracted. In TMT, the typical value is approximated by the geometric disorder average average\cite{Dobrosavljevic2003,Byczuk2005,Byczuk2009,Aguiar2009,Byczuk2010,Dobrosavljevic2010}
\begin{equation}
\langle \rho (\omega) \rangle_{\mbox{\tiny geo-dis}} =  \exp \int d\epsilon \; p_{\epsilon}(\epsilon) \, \ln \rho_{\epsilon}(\omega).\label{eq_geo-dis_av}
\end{equation}
Since statistical DMFT gives access to the full PDF of the LDOS, we will consider the geometric average of the PDF   
\begin{equation}
\langle \rho (\omega) \rangle_{\mbox{\tiny geo}} =  \exp \int d\rho'(\omega) \; p[\rho'(\omega)] \, \ln \rho'(\omega) \label{eq_geo_av}
\end{equation}
only for comparison purposes. In the following, we will also use the cumulative PDFs 
\begin{equation}
P[\rho (\omega)] =  \int\limits_0^{\rho(\omega)}  d\rho' (\omega)\; p[\rho'(\omega)] \;. \label{eq_cumulative}
\end{equation}

To the end of this section we shortly introduce the impurity solver that is used to solve the impurity part of the DMFT equations. Since we are dealing with a large number of single sites, we have to be restricted to impurity solvers with a feasible amount of computational time. In this work we employ the MPT,\cite{Kajueter1996,Potthof1997} which properly reproduces the non-interacting limit and the atomic limit. MPT is the generalization of the iterated perturbation theory (IPT)\cite{MullerHartmann1989,Zhang1993,Georges1996} to systems with arbitrary fillings. IPT 
qualitatively describes the Mott-Hubbard metal-insulator transition at a finite critical interaction strength $U_c$\cite{Zhang1993,Georges1996} by calculating the self-energy to the second order in $U$ in the non-renormalized perturbation expansion. Using this impurity solver, Green's function ensemble sizes of the order $N\sim10^3$ are computationally feasible within a parallelized code run on a cluster.

Within MPT the local self-energy is given by\cite{Kajueter1996,Potthof1997}
\begin{equation}
\Sigma_i(\omega) = U \langle n_i \rangle + \frac{a_i \Sigma_i^{(2)}(\omega)}{1 - b_i \Sigma_i^{(2)}(\omega)} \,, \label{eq_M6}
\end{equation}
where $\Sigma_i^{(2)}(\omega)$ is the second order perturbation contribution to the self-energy.\cite{MullerHartmann1989} The perturbation expansion uses the non-renormalized Hartree-Fock Green's functions\cite{Potthof1997} 
\begin{equation}
G_i^{\mbox{\tiny{HF}}} (\omega) =  \frac{1}{\omega+\tilde{\mu}_i-\epsilon_i-U \langle n_i\rangle-\Gamma_i(\omega)+ i\eta}  
\end{equation}
as propagators. The parameter $\tilde{\mu}_i$ is fixed by requiring $\langle n_i \rangle = \langle n_i \rangle^{\mbox{\tiny{HF}}} $.\cite{Potthof1997} The parameters $a_i$ and $b_i$ are determined by the requirement that the spectral moments 
\begin{equation}
M_i^{(m)} = \int\limits_{-\infty}^{\infty} d\omega \; \omega^m \rho_i (\omega) 
\end{equation}
reproduce an equation of motion analysis up to $m=3$ and that the atomic limit is recovered correctly.\cite{Potthof1997} The corresponding self-consistency equations that accomplished the requirement were worked out by M. Potthoff \textit{et al.}\cite{Potthof1997}

\section{Results}\label{results}

In our calculations, we consider the half-filled case, i.e. band filling $\nu=\frac{1}{N}\sum_{i\sigma}\langle  n_{i\sigma}\rangle =1$, which is accomplished by setting $\mu=U/2$ for box disorder. We work in energy units of the non-interacting bandwidth $W_0=4t\sqrt{K}=1$. Since we used the approximate equation (10), the DOS of non-disorder system is always semielliptic for each $K$. 

In order to investigate the physics of strongly correlated electrons in disordered lattices we need to define all relevant phases. The correlation-induced \emph{Mott insulator} is incompressible and exhibits an energy gap in the spectrum, which is proportional to the interaction strength $U$. A gapped spectrum in turn corresponds to a vanishing arithmetic average of the LDOS at the Fermi level, i.e. $\langle \rho(\omega=0)\rangle_{\mbox{\tiny arith}}=0$. An insulating phase that originates from the randomness is the well-known \emph{Anderson insulator}\cite{Anderson1958} for zero interaction. This phase is compressible and its spectrum is dense point-like, which can be attributed to an absence of diffusion.\cite{Economou1972} In the interacting case, we define a state to be localized or extended, if the spectrum of the single-particle Green's function is a dense point-like or continuous respectively, which is a natural generalization of the localization concept to interacting systems. Consequently, the \emph{Anderson-Mott insulator} is defined by localized single-particle excitations at the Fermi level. So far it is not established on a mathematical level of rigor, that localized single-particle excitations correspond to a vanishing dc conductivity, but strong evidence was recently given in case of weak interactions.\cite{Basko2006} How we numerically detect localized states within the statistical DMFT will be explained in the following subsection \ref{sec_localization}. Finally, the \emph{paramagnetic metal} is compressible and therefore has a non-vanishing DOS $\langle \rho(\omega=0)\rangle_{\mbox{\tiny arith}}$ at the Fermi level and the corresponding states are extended.

\subsection{Anderson transition in the non-interacting case}\label{sec_localization}

\begin{figure}[t]
\centering
\includegraphics[width=0.48\textwidth]{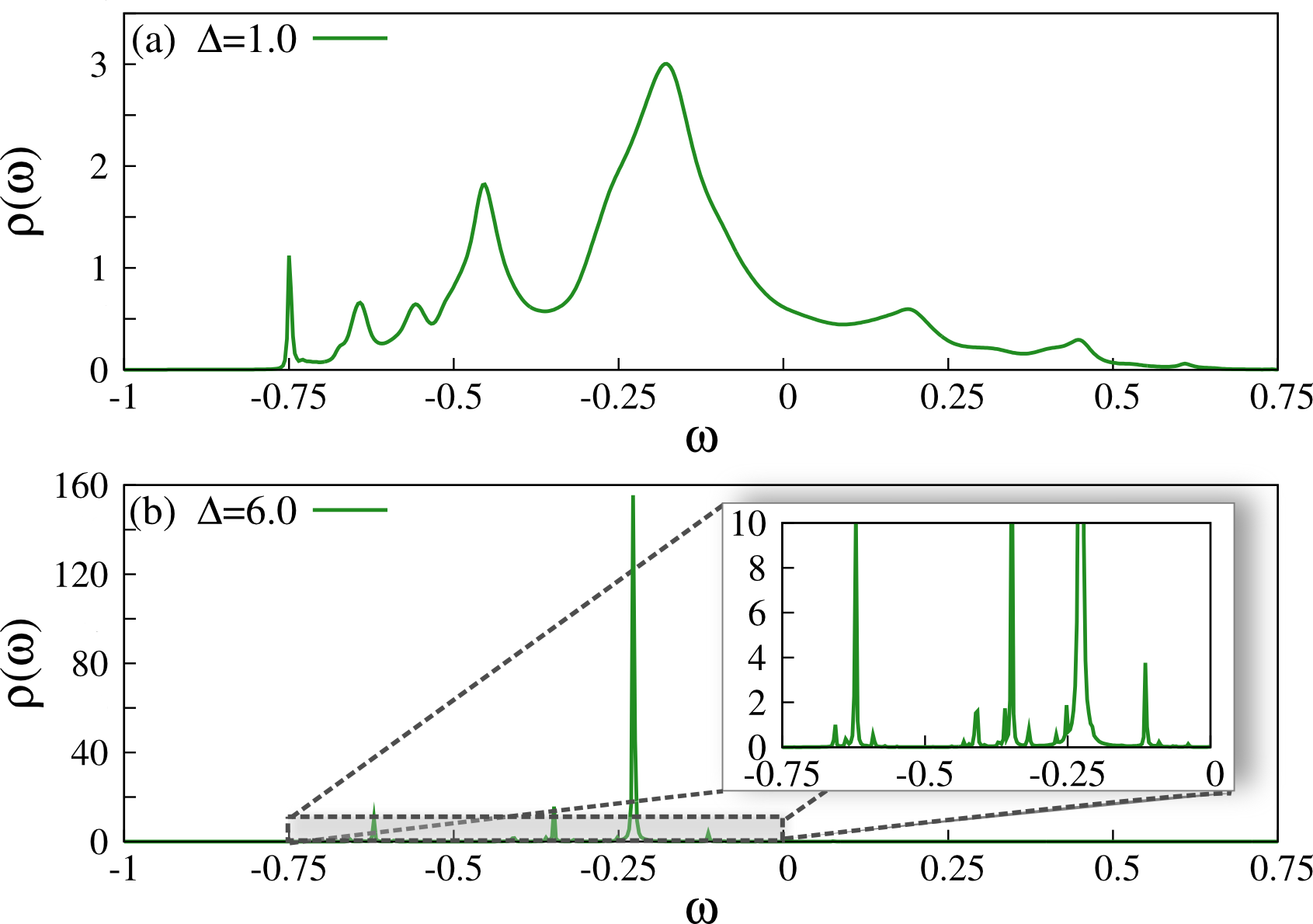}  
\caption{(Color online) Local density of states $\rho(\omega)$ of one random sample out of the ensemble for two different box disorder strengths $\Delta$: (a) $\Delta=1$ and (b) $\Delta=6$. Parameters are $\eta=10^{-3}$, $U=0$, $W_0=1$, and $K=6$.}
\label{fig_spectra_extended_localised}
\end{figure}

Localized states are defined by the absence of diffusion:\cite{Anderson1958} The state of a propagating particle initially localized at one single lattice site exhibits a finite return probability in the thermodynamic limit. It was shown that the finite return probability implies a dense distribution of poles on the real axis of the single-particle Green's function in contrast to extended states, which are given by a branch cut on the real axis of the single-particle Green's function.\cite{Economou1972} 
This difference in the analytic properties of the single-particle Green's function can be utilized for a numerical distinction between localized and extended states within the statistical DMFT and its non-interacting limit, i.e. the local distribution approach.\cite{Abou-Chacra1973,Alvermann2005,Alvermann2006} Extended and localized states are characterized by different behavior of the PDF $p[\rho(\omega)]$ in the limit of vanishing broadening $\eta\rightarrow 0$,\cite{Semmler2010,Semmler2010b,Alvermann2005} as is explained subsequently in more detail. 

In a finite lattice the local one-particle Green's function at site $0$ is given by \cite{Licciardello1975}
\begin{equation}
G_{00}(\omega)= \sum\limits_n \frac{f_n}{\omega-E_n} \;. \label{eq_ef_representation_GF}
\end{equation}
Here, $f_n= \langle0|\psi_n \rangle \langle\psi_n|0 \rangle$ denotes the overlap of the eigenstate $n$ with the Wannier function on site $0$, $|\psi_n\rangle$ denotes the complete set of single particle energy eigenstates  with eigenenergies $E_n$. It was shown that the return probability is determined by $p_{0\rightarrow 0}(t\rightarrow \infty)=\sum_n f_n^2$.\cite{Licciardello1975} In case of an extended eigenstate, the residue $f_n$ is proportional to the inverse number of occupied lattice sites $N^{-1}$. Consequently, the return probability approaches zero in the thermodynamic limit. 

In case of an exponentially localized state, the residues are finite even in the thermodynamic limit. If these values were ordered, they would decrease exponentially and only some of them would contribute significantly to the sum (\ref{eq_ef_representation_GF}). In particular, if we introduce a small coupling $\eta$ to a dissipative bath, which regularizes the singularities, and consider a contour which encloses a small energy interval $\delta E$, then the most probable value of the sum of the residues of poles enclosed by this contour will decrease exponentially proportional to the ratio $\Delta/\delta E$. From here it follows, that the most probable value of the imaginary part of the Green's function is proportional to $\eta$.\cite{Thouless1970} Hence, the LDOS will be highly fragmented in case of localized eigenstates, characterized by dominating well-separated resonances. In contrast, the arithmetically averaged spectral function, i.e. the DOS of the system, does not exhibit this high fragmentation, since the spectral weight must be located somewhere in the lattice for every energy. 

This theoretical reasoning can be verified within our numerical investigation. As explained in the method section we determine the PDF of the local Green's function by calculating the local Green's function for each sample of the ensemble. Each of them is given by a random on-site energy, a local self-energy and a hybridization function which is randomly sampled by $K$ local Green's functions out of the ensemble according to equation \ref{eq_hybrid}.
Having calculated all local Green's functions, we can investigate the local spectrum of one single sample out of the ensemble. Two example spectra are displayed in Fig.~\ref{fig_spectra_extended_localised}. Panel (a) shows a typical spectrum for small disorder strengths, namely $\Delta=1$. The spectrum is smooth and furthermore broadened in comparison to $W_0$ due to the disorder. In contrast, the spectrum at higher disorder strength $\Delta=6.0$, shown in panel (b), is highly fragmented and consists of various delta peaks, which are broadened by the artificial value $\eta$. This can be seen more clearly in the inset. In fact, this is exactly what we expect: Extended states of the system are characterized by a branch cut on the real axis of the local Green's function, whereas localized states are given by poles.
      
\begin{figure}[t]
\centering
\includegraphics[width=0.48\textwidth]{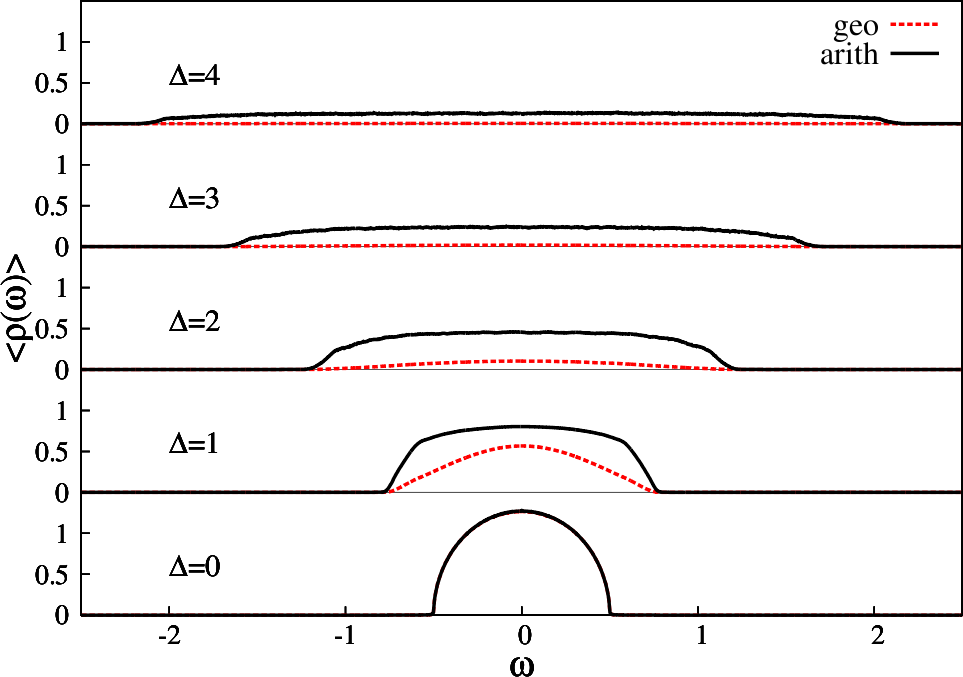}  
\caption{(Color online) Evolution of the arithmetically and geometrically averaged spectral function of the non-interacting system with increasing disorder strength $\Delta$. Parameters are $\eta=10^{-3}$, $U=0$,  $W_0=1$, and $K=6$.}
\label{fig_spectra_no_ww}
\end{figure}

So far, the spectrum of a single sample was considered, but it is also interesting to study the distribution of spectra of the whole ensemble. The disorder causes a broadening of the spectrum as can be clearly seen from Fig.~\ref{fig_spectra_no_ww}, where the geometrically averaged (\ref{eq_geo_av}) and arithmetically averaged spectral functions (\ref{eq_arith_av}) are plotted for selected values of the disorder strength and zero interaction. The broadening is naturally accompanied by a drop-off of the geometric average as well as the arithmetic average. This is because the spectral weight is distributed over a bigger range of energies with increasing disorder strength. As an important feature we note the much stronger decline of the geometric average to zero, as found previously within TMT.\cite{Byczuk2010,Dobrosavljevic2010} Since the arithmetically and geometrically averaged spectral function are plotted as a function of the energy, we are also able to note that the decline is not uniform, but rather strongly depends on the energy. For instance, the geometric average of the LDOS remains clearly finite in the band center for $\Delta=2$, whereas it is close to zero for the states in the outer parts of the spectrum.

\begin{figure}[t]
\centering
\includegraphics[width=0.48\textwidth]{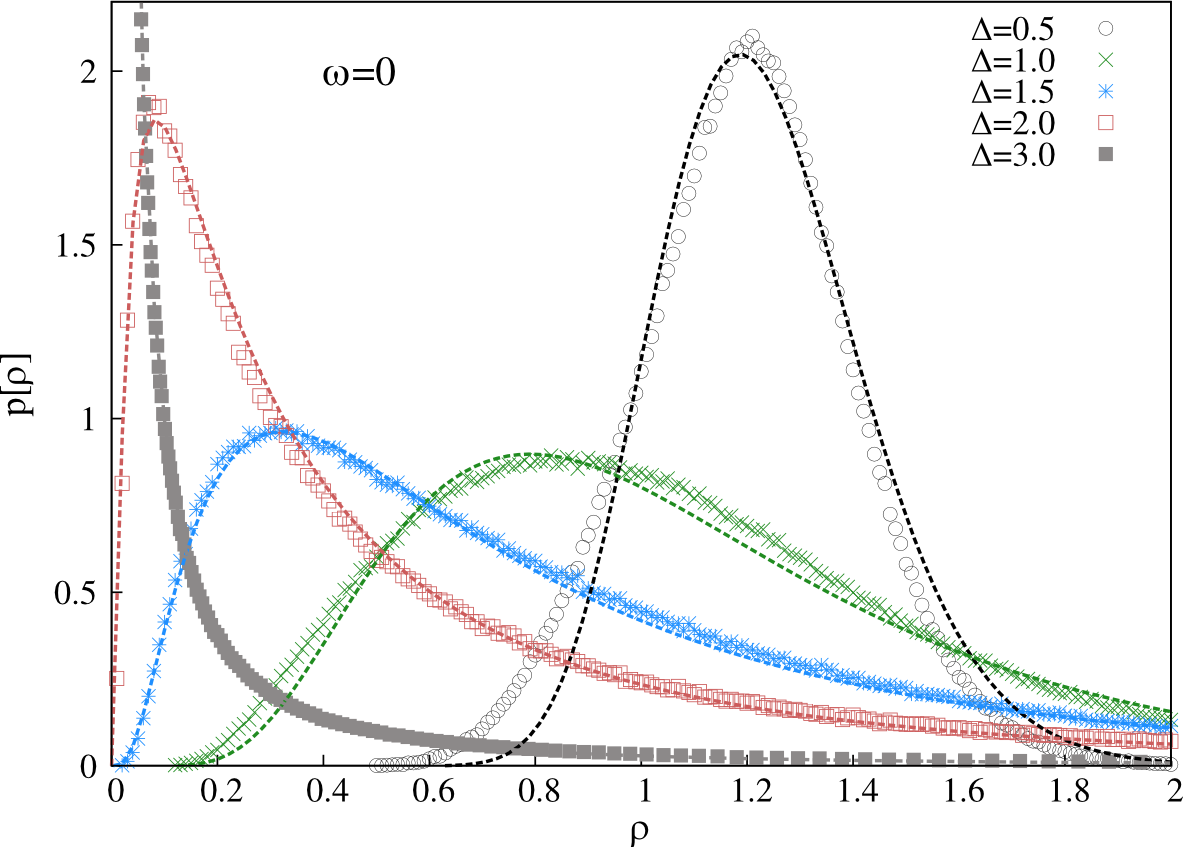}  
\caption{(Color online) Evolution of the PDF of the LDOS at the Fermi level $p[\rho(0)]$ for increasing disorder strength $\Delta$. The dashed lines correspond to least square fits with a log-normal function. Parameters are $W_0=1$, $\eta=10^{-6}$ and connectivity $K=6$.}
\label{fig_histo_development_with_D}
\end{figure}

Next we discuss how the overall PDF of the LDOS is affected by disorder. In Fig.~\ref{fig_histo_development_with_D} the PDF of the LDOS is displayed for several disorder strengths. The initial delta function for the homogeneous system located at $\rho=1.26$ (not displayed) is broadened by a small amount of disorder ($\Delta=0.5$) due to the disorder-induced fluctuations. By increasing the disorder strength further to $\Delta=1,1.5,2$ and $3$ the PDF spreads more and more, long tails develop and more weight is shifted to smaller values. This becomes evident, if we focus on the most probable value of the PDF, also called the \textit{typical value}, which approaches zero with increasing disorder strength. The dashed lines in Fig.~\ref{fig_histo_development_with_D} correspond to least square fits with theoretically predicted log-normal distributions\cite{Mirlin1996,Mirlin2000}
\begin{equation}
p_{\mbox{\tiny{ln}}}(\rho) = \frac{1}{\sqrt{2\pi \sigma^2}\rho} \exp \left(-\frac{(\ln\rho-\mu)^2}{2 \sigma^2} \right)  
\end{equation}
 with fitting parameters $\mu$ and $\sigma$. Obviously, the approximation becomes better for stronger disorder, in agreement with recent high accuracy kernel polynomial method calculations for various Bravais lattices in the non-interacting system.\cite{Schubert2010}  

\begin{figure}[t]
\centering
\includegraphics[width=0.48\textwidth]{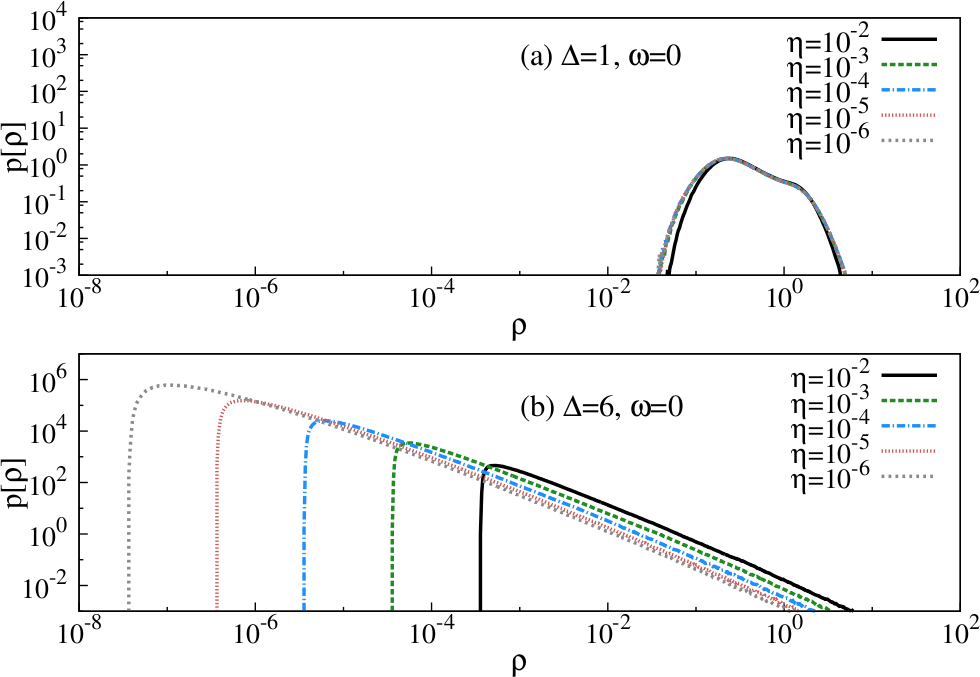}  
\caption{(Color online) Evolution of the PDF of the LDOS at the Fermi level $p[\rho(0)]$ for decreasing broadening $\eta$ at  two different disorder strengths $\Delta$: (a) $\Delta=1$ and (b) $\Delta=6$. Parameters are $K=6$, $U=0$, and $W_0=1$.}
\label{fig_localized_extended}
\end{figure}

Now we discuss how the broadening $\eta$ is used for a distinction between localized and extended states. In the localized phase the Green's function is given by a distribution of poles. Hence, an arbitrarily chosen frequency $\omega$ lies with probability one between the poles, resulting in a value of the LDOS equal to zero. If it is exactly a $\delta$-peak, the resulting value would be infinity. However, the probability for this event is zero. The artificially introduced coupling to a dissipative bath via the small factor $\eta$ broadens the $\delta$-peaks to Lorentzians with a width proportional to $\eta$ and generates a finite probability to obtain a finite value of the LDOS. Moreover, let us consider an energy in the vicinity of a pole present in the considered sample. The corresponding value of the LDOS is proportional to $\eta$, since the LDOS is given by the imaginary part of the single-particle Green's function. Now the observed features of the PDF of the LDOS in Fig. \ref{fig_localized_extended} can be understood: In a system where the local spectra are highly fragmented due to the disorder, most ensemble samples will exhibit a nearly zero ($\sim \eta$) value of the LDOS for a given frequency $\omega$. On the other hand, a small fraction will contribute high values of the LDOS stemming from the broadened $\delta$-peaks, constituting the long tails of the PDF. Consequently, the maximum of the PDF of the LDOS of an extended state saturates at a finite value for $\eta\rightarrow 0$, but an increasing amount of the PDF's weight shifts to zero in the case of localized states.

\begin{figure}[t]
\centering
\includegraphics[width=0.48\textwidth]{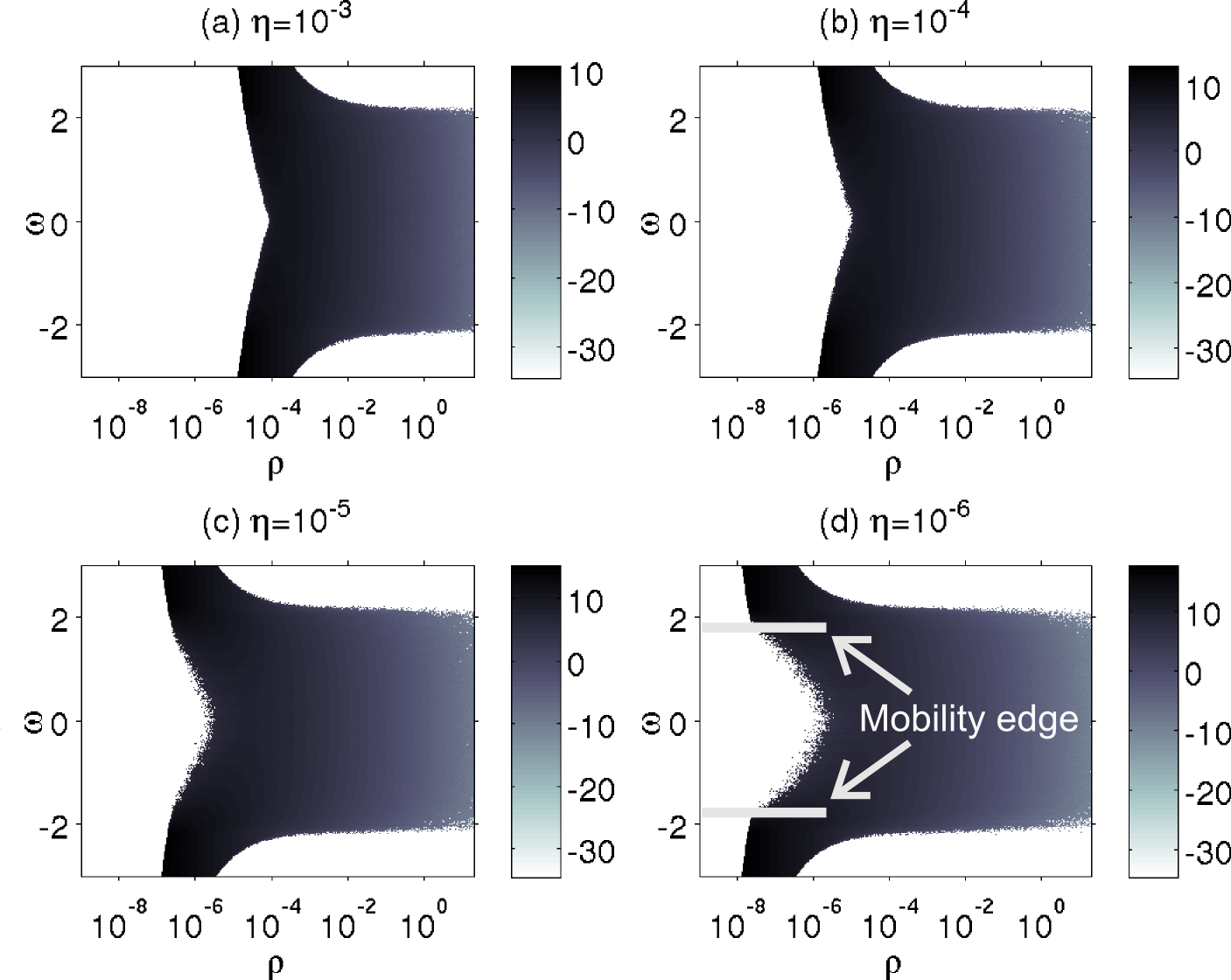}  
\caption{(Color online) Natural logarithm of the PDF $p[\rho(\omega)]$ for fixed  disorder strength $\Delta=4$ and different values of $\eta$: (a) $\eta=10^{-3}$, (b) $\eta=10^{-4}$, (c) $\eta=10^{-5}$ and (d) $\eta=10^{-6}$. Parameters are $K=6$, $U=0$, and $W_0=1$.}
\label{fig_mobility_edge}
\end{figure}

This allows for a numerical distinction between localized and extended states. Clearly, the PDF of the LDOS of an extended state will not be affected upon lowering of $\eta$ below a certain value. In contrast, in the localized phase, the PDF will be affected dramatically: With decreasing broadening $\eta$ an increasing amount of the PDF's weight is shifted to smaller values,  and the most probable value will shift to zero. Fig.~\ref{fig_localized_extended} displays the evolution of the PDF of the LDOS at frequency $\omega=0$ on a log-log scale when the broadening is decreased. In panel (a) the disorder strength $\Delta$ is equal to one and the PDF is found to saturate. We conclude that the state is extended. For $\Delta=6$ in panel (b) the PDF exhibits the above described behavior. We conclude that the state is localized.

It should be shortly mentioned here that the PDFs in the localized phase are not well described by log-normal functions. In fact, the shape of the PDFs results from a collection of narrow Lorentzians. The width of these Lorentzians is determined by the broadening $\eta$ and so is the shape of the PDF. Therefore, the log-normal distributions can only be used as suitable approximations for extended states due to the finite broadening.   

When the analysis is performed with a complete frequency resolution, the mobility edges of the system can be identified. For example this is done in Fig.~\ref{fig_mobility_edge}, where the natural logarithm of the PDF $p[\rho(\omega)]$ is shown for fixed disorder strength $\Delta=4$ in the $\omega$-$\rho$-plane. The broadening factor $\eta$ is lowered from $10^{-3}$ in panel (a) to $10^{-6}$ in panel (d). When the broadening is decreased, we notice clearly a low-energy core of the band, where the PDFs are unaffected upon further lowering of the broadening. On the other hand, outside of this core the PDF shows the behavior for localized states. The separating energies represent the mobility edges we are looking for. It is important to note that the resolution of this procedure is determined by the lowest possible value of $\eta$, which is basically determined by the finite bath size. In any case, the procedure will give the lower bound of the critical disorder strength for the transition from extended to localized states, since an extended state will never be mistaken as a localized state. This question of resolution was examined in more detail in Ref.~\onlinecite{Alvermann2005}. 

When the mobility edges cross the Fermi level in the band center all states become localized, corresponding to the Anderson insulator. Since localized states do not contribute to the DC conductivity, a metal-insulator transition is obtained, caused by the disorder. The existence of this transition on the Bethe lattice was analytically shown \cite{Abou-Chacra1973} and recently checked numerically.\cite{Alvermann2006} 

\begin{figure}[t]
\centering
\includegraphics[width=0.48\textwidth]{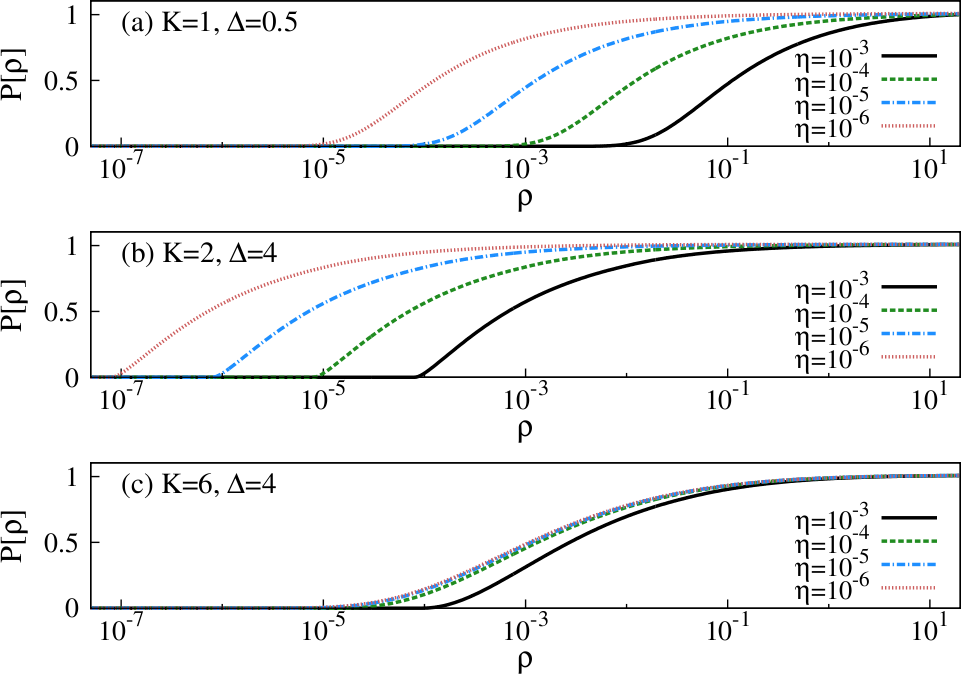}  
\caption{(Color online) Evolution of the cumulative PDFs at the Fermi level $P[\rho(0)]$ of the non-interacting system with decreasing $\eta$ in (a) for $K=1$ and $\Delta=0.5$, (b) for $K=2$ and $\Delta=4$, and (c) for $K=6$ and $\Delta=4$. Parameters are $U=0$ and $W_0=1$.}
\label{fig_Psum_vgl_K}
\end{figure}

Compared to CPA and TMT, the connectivity $K$ represents an additional free parameter within the statistical DMFT. It is well established that no localization occurs in the limit of infinite spatial dimensions, or for infinite connectivity, respectively. On the other hand, it is known from scaling theory that in one dimension and in two dimensions -- i.e. for small connectivities -- already arbitrarily small disorder suffices to localize all states in the absence of interactions.\cite{Abrahams1979} Hence, this strong dependence of localization on the connectivity should be captured by statistical DMFT. We expect that for higher connectivity the tendency towards delocalization is enhanced. Let us, for instance, consider the evolution of the cumulative PDF of the LDOS at the Fermi level defined in equation (\ref{eq_cumulative}) when the broadening $\eta$ approaches zero for several parameter sets, as given in Fig.~\ref{fig_Psum_vgl_K}. In panel (a) we show a small disorder strength $\Delta=0.5$ for connectivity $K=1$, i.e. a linear chain. We note that the $\omega=0$ state is localized. For $K=2$ (not displayed) our analysis shows that the state is extended and higher disorder strengths are needed to localize the state at the Fermi level, as can be seen for example for $\Delta=4$ in panel (b). Finally, in panel (c) we investigate the connectivity $K=6$ for the same disorder strength. In the former case the state at $\omega=0$ is localized, whereas in the latter case with a larger connectivity the state is extended. 

\subsection{PDFs of the LDOS for finite interactions}

\begin{figure}[t]
\centering
\includegraphics[width=0.48\textwidth]{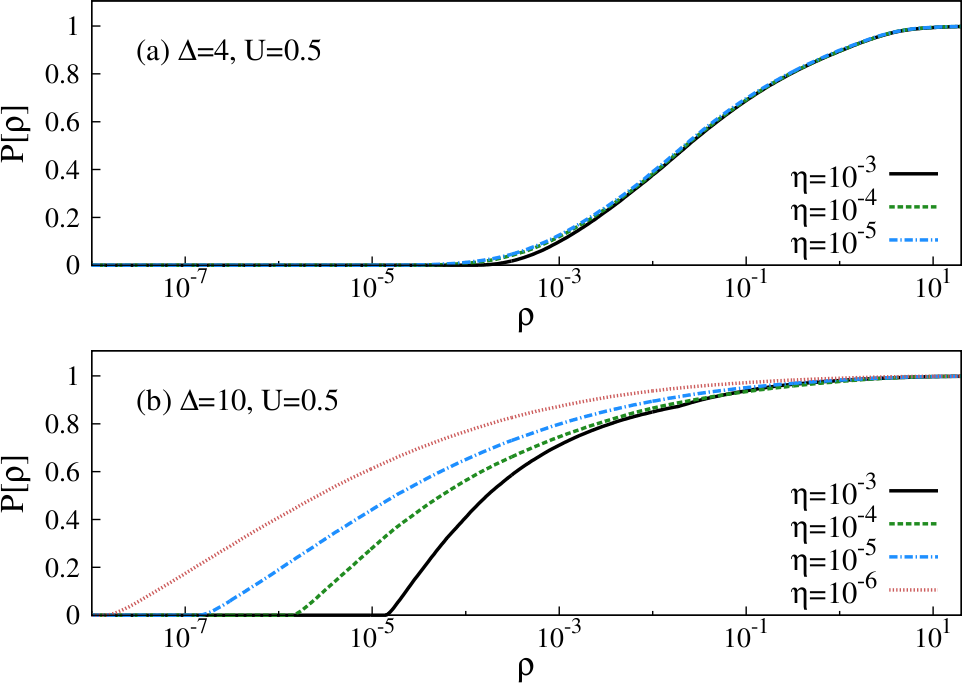}  
\caption{(Color online) Evolution of the cumulative PDFs at the Fermi level $P[\rho(0)]$ with decreasing $\eta$ (a) for $\Delta=4$ and (b) for $\Delta=10$ at fixed interaction strength $U=0.5$. Parameters are $K=3$, $W_0=1$, and $\mu=U/2$.}
\label{fig_loc_with_U}
\end{figure}

In this subsection we will address the question of disorder-induced localization in the presence of interactions. The evolution of the cumulative PDF of the LDOS with decreasing broadening is plotted in Fig.~\ref{fig_loc_with_U} for the weakly interacting case $U=0.5$. In panel (a) the disorder strength is given by $\Delta=4$ and in panel (b) $\Delta=10$, corresponding to the strongly disordered regime. We note that the PDF saturates for $\Delta=4$ when the broadening approaches zero, corresponding to an extended state. In contrast, for very strong disorder strength $\Delta=10$ the PDF does not saturate. Here, the state is localized within the resolution given by the size of our stochastic Green's function bath. By comparison to Fig.~\ref{fig_Psum_vgl_K}, we note that the shape of the PDFs of the interacting system differs from those in the non-interacting case. In particular, the slope of the cumulative PDFs varies with the broadening, which indicates that significant weight of the PDF is distributed over several orders of magnitude of the LDOS.

\begin{figure}[t]
\centering
\includegraphics[width=0.48\textwidth]{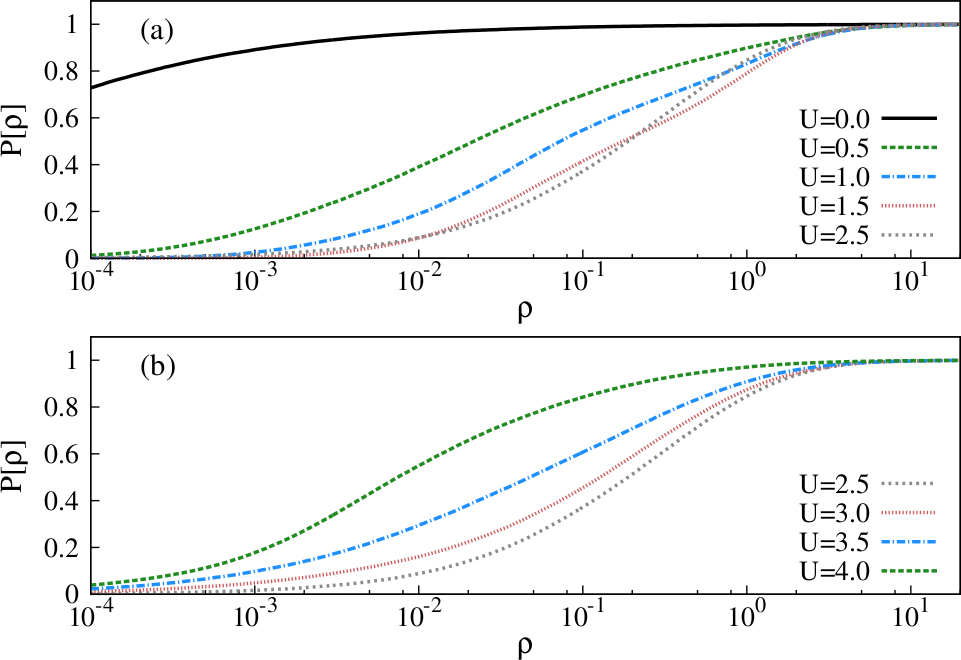}  
\caption{(Color online) Evolution of the cumulative PDF at the Fermi level $P[\rho(0)]$ with increasing interaction $U$ for fixed disorder strength $\Delta=4$. In panel (a) the PDF of the non-interacting system is compared to the PDFs of the interacting system from weak interactions ($U=0.5$) to intermediate interactions ($U=1,1.5,2$). In panel (b) the interaction strength $U$ is further increased stepwise to the strongly interacting regime ($U=3,3.5,4$). Parameters are $K=3$, $\eta=10^{-5}$, $W_0=1$, and $\mu=U/2$.}
\label{fig_pdf_vary_U}
\end{figure}

The statistical DMFT allows for studying how the PDF of the LDOS evolves from the non-interacting case and weakly interacting case to the regime of strong correlations. In Fig.~\ref{fig_pdf_vary_U} the PDFs are plotted for several selected values of the interaction and fixed disorder strength $\Delta=4$ and connectivity $K=3$. In panel (a) weak interactions ($U=0.5$) are introduced and stepwise increased to the intermediate regime ($U=2.5$). In comparison to the non-interacting case (corresponding to a localized state as we will see clearly in the next section), we note a striking redistribution already for weak interactions. The PDF is still extended over many decades of values of the LDOS, but a big fraction of the PDF is shifted to much higher values of the LDOS compared to the non-interacting case. We will understand this severe modification later. Further increase of the interaction strength to $U=1.5$ systematically fortifies this redistribution. More and more sites exhibit comparable values of the LDOS, as can be seen from the larger slope of the cumulative PDF. This behavior contradicts a strong fragmentation of the single-particle excitation spectrum. Moreover, we note that the cumulative PDF gains most of its weight in two regions of a big slope for $U=1.5$. This feature erodes for bigger interaction strengths as can be seen from the cumulative PDF for $U=2.5$. In panel (b) the interaction strength is stepwise increased further to $U=4$. In contrast to the behavior found in panel (a) it can be clearly seen that the opposite behavior is obtained in that case. The PDFs shifts again to smaller and smaller values of the LDOS. Obviously 
correlations do not monotonically influence the PDF of the LDOS, but rather lead to two different regimes. We expect this behavior to have a strong influence on the conduction properties of the system, and that it should  also be mirrored in other observables. 

\begin{figure}[t]
\centering
\includegraphics[width=0.48\textwidth]{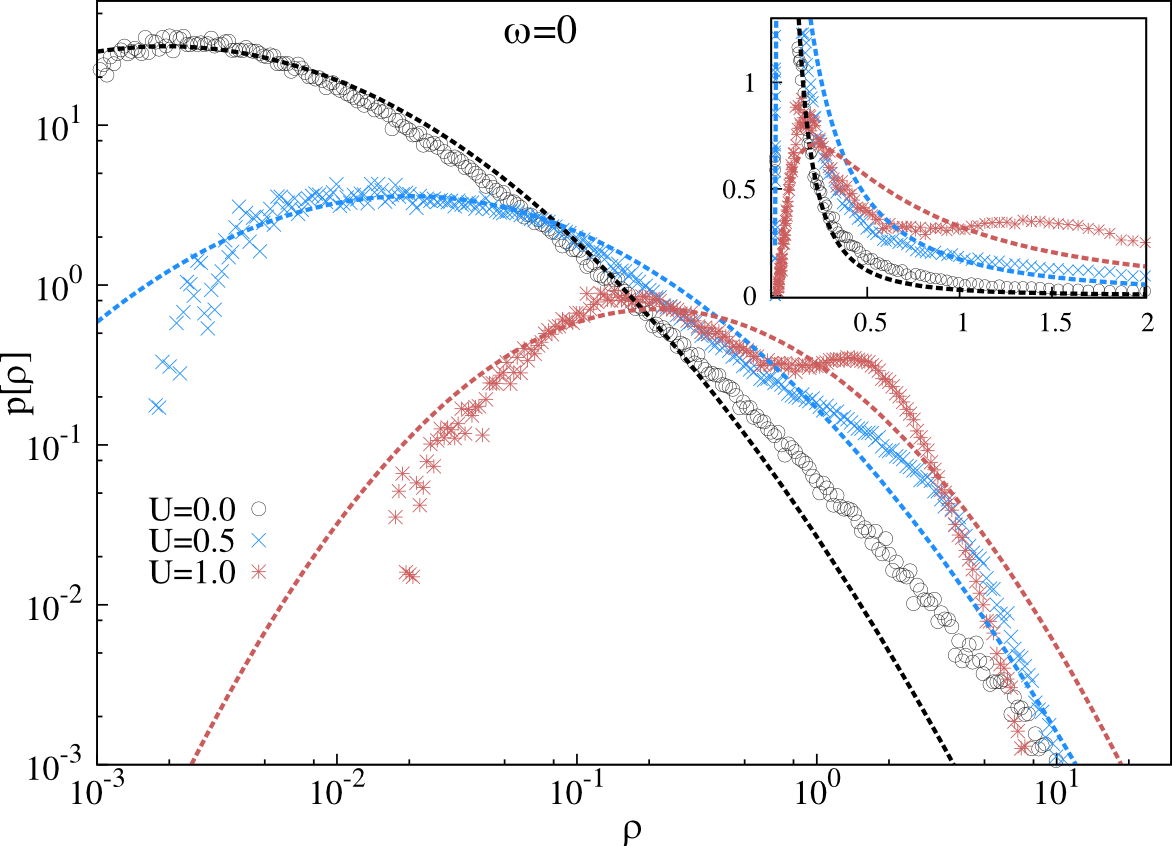}  
\caption{(Color online) Evolution of the PDF of the LDOS at the Fermi level $p[\rho(\omega=0)]$ for fixed disorder strength $\Delta=2$ and increasing interaction: $U=0,0.5,1$. The dashed lines correspond to least square fits with a log-normal function. Parameters are $K=3$, $\eta=10^{-3}$, $W_0=1$ and $\mu=U/2$.}
\label{fig_ln_fit_finiteU}
\end{figure}

To understand this behavior the PDF is compared to least squares fits of log-normal distributions for $\Delta=2$ in Fig.~\ref{fig_ln_fit_finiteU}. At zero interaction the log-normal distribution represents a good approximation of the PDF, but for increasing interaction strengths $U=0.5$ and $U=1.0$ strong modifications are observed. The weight of the PDF is shifted to higher values of the LDOS and an additional interaction-induced peak is observed. The latter becomes particularly apparent on a linear scale as shown in the inset. From these results it is obvious that the log-normal distribution is not sufficient to approximate the PDF of the LDOS in the interacting case. It is interesting to note that the formation of a second peak is reversed upon further increase of the interaction strength as we learned from the discussion of Fig.~\ref{fig_pdf_vary_U}. 

\begin{figure}[b]
\centering
\includegraphics[width=0.48\textwidth]{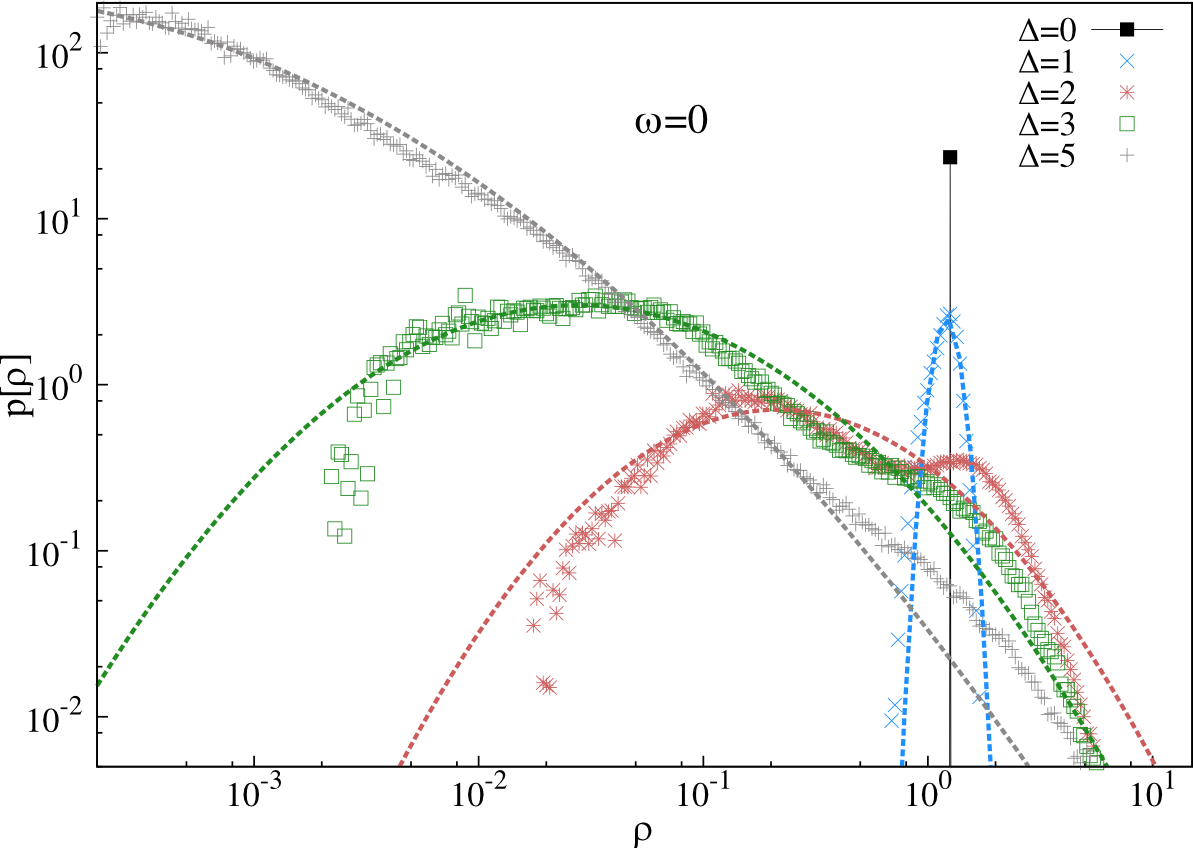}  
\caption{(Color online) Evolution of the PDF of the LDOS at the Fermi level $p[\rho(\omega=0)]$ for fixed interaction strength $U=1$ with increasing disorder strength: $\Delta=0,1,2,3,5$. The dashed lines correspond to least square fits with a log-normal function. Parameters are $K=3$, $\eta=10^{-3}$, $W_0=1$, and $\mu=U/2$.}
\label{fig_ln_fit_finiteU_vglD}
\end{figure}
Next we discuss the evolution of the PDF with increasing disorder and for fixed interaction strength $U=1$ as plotted in Fig.~\ref{fig_ln_fit_finiteU_vglD}. In the homogeneous system the PDF of the LDOS is given by a delta-function at the value $\rho=1.26$, which corresponds to the value of the non-interacting system due to the Luttinger theorem. Finite disorder strength ($\Delta=1$) broadens the PDF similar to the behavior found in the non-interacting case as shown in Fig.~\ref{fig_histo_development_with_D}. However, we clearly recognize the two-peak structure of the PDF for $\Delta=2$ and $\Delta=3$. This structure is also observable for $\Delta=5$, but a suitable approximation via the log-normal distribution is clearly restored in the strongly disordered regime.

\subsection{Paramagnetic ground state phase diagram}\label{sec_phase_diagram}

\begin{figure}[t]
\centering
\includegraphics[width=0.48\textwidth]{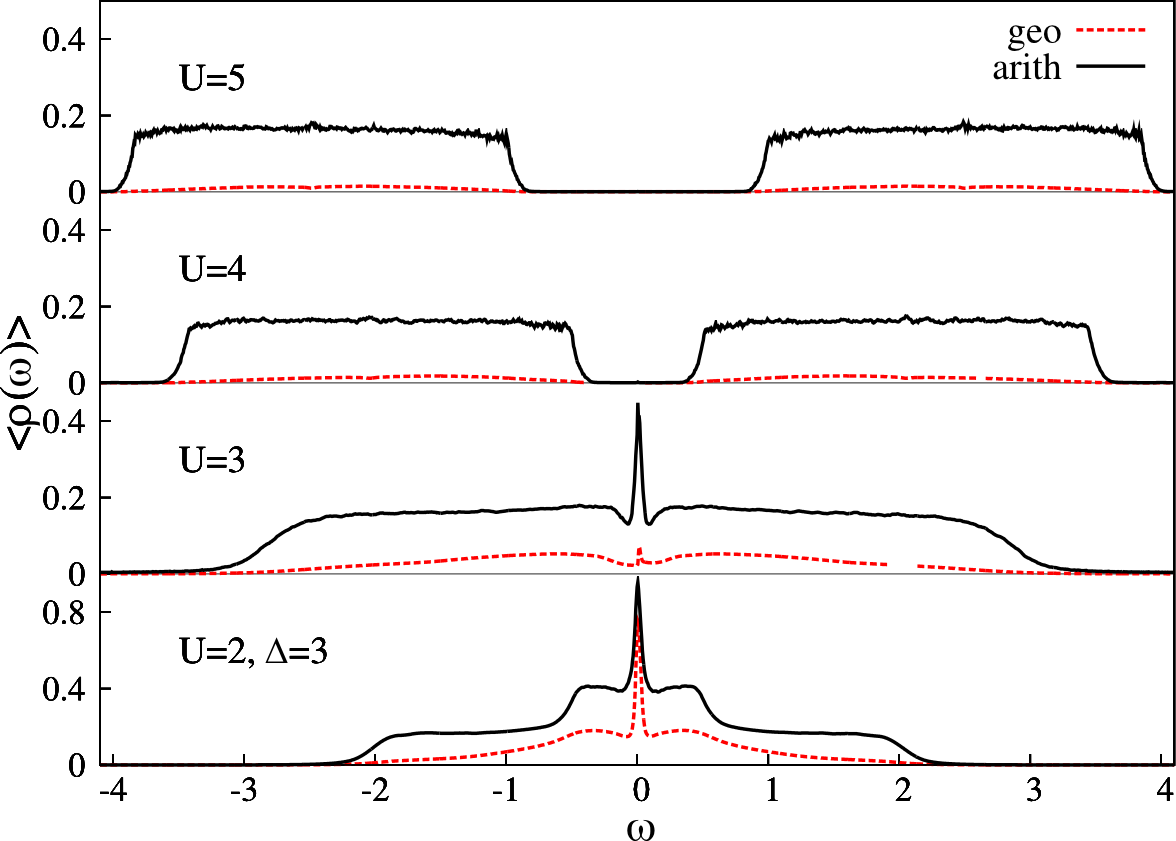}  
\caption{(Color online) Evolution of the arithmetically and geometrically averaged spectral function with increasing interaction ($U=2,3,4,5$) for fixed disorder strength $\Delta=3$. Parameters are $\eta=10^{-3}$, $\mu=U/2$, $W_0=1$, and $K=3$.}
\label{fig_box_mott_D3}
\end{figure}

An important feature of the system considered here is given by the Mott transition in the homogeneous system. By construction the statistical DMFT reduces to DMFT in the homogeneous case.  Since box disorder exhibits a bounded range of accessible on-site energies, it is reasonable to expect the Mott transition to be also present in the disordered case at least for moderate disorder strengths. An example of such a transition is given in Fig.~\ref{fig_box_mott_D3}, in which the geometrically and arithmetically averaged spectral functions are displayed for fixed disorder strength $\Delta=3$ and increasing interaction. For $U=2$ we recognize a spectral function corresponding to a disordered metal, as at the same time it was checked that the states at the Fermi level are extended. This means that the system exhibits no excitation gap. When the interaction strength is increased ($U=3$), the spectrum becomes more and more dominated by incoherent excitations. Finally, the spectra belonging to $U=4$ and $U=5$ evidently correspond to Mott insulating states, since $\langle\rho(0)\rangle_{\mbox{\tiny{arith}}}=0$. Obviously, a Mott transition takes place between $U=3$ and $U=4$.

Moreover, it is interesting to note that we find no evidence for a zero-bias anomaly as can be seen from the spectra. In contrast, recent exact diagonalization investigations of the Anderson-Hubbard Hamiltonian within the Hartree-Fock approximation by Shinaoka and Imada\cite{Shinaoka2009,Shinaoka2009b} established the existence of a soft Hubbard gap. The physical origin of the zero-bias anomaly was analyzed in more detail by Chen {\it et al.}\cite{Chen2011} pointing out the importance of non-local correlations which are not included in our theory. Please note that in our recent statistical DMFT investigation of fermions in speckle disordered optical lattices\cite{Semmler2010b} we compared results for the Bethe lattice as presented here with results for a finite-size square lattice. We found evidence for the existence of a zero-bias anomaly on the square lattice but not on the Bethe lattice. 

It is of fundamental interest how the Anderson transition of the non-interacting system is modified in the presence of  interactions and how the Mott transition of the homogeneous system is affected by disorder. The nature of the Anderson-Mott transition has been investigated in detail within TMT\cite{Aguiar2006,Aguiar2009,Dobrosavljevic2010} and within statistical DMFT.\cite{Dobrosavljevic1997} In the latter, also a Griffiths phase was revealed as precursor for the Anderson-Mott transition. However, the extent of the metallic phase has not been systematically discussed within statistical DMFT so far, but only within TMT, which neglects spatial fluctuations.\cite{Aguiar2009,Byczuk2005}  

\begin{figure}[b]
\centering
\includegraphics[width=0.48\textwidth]{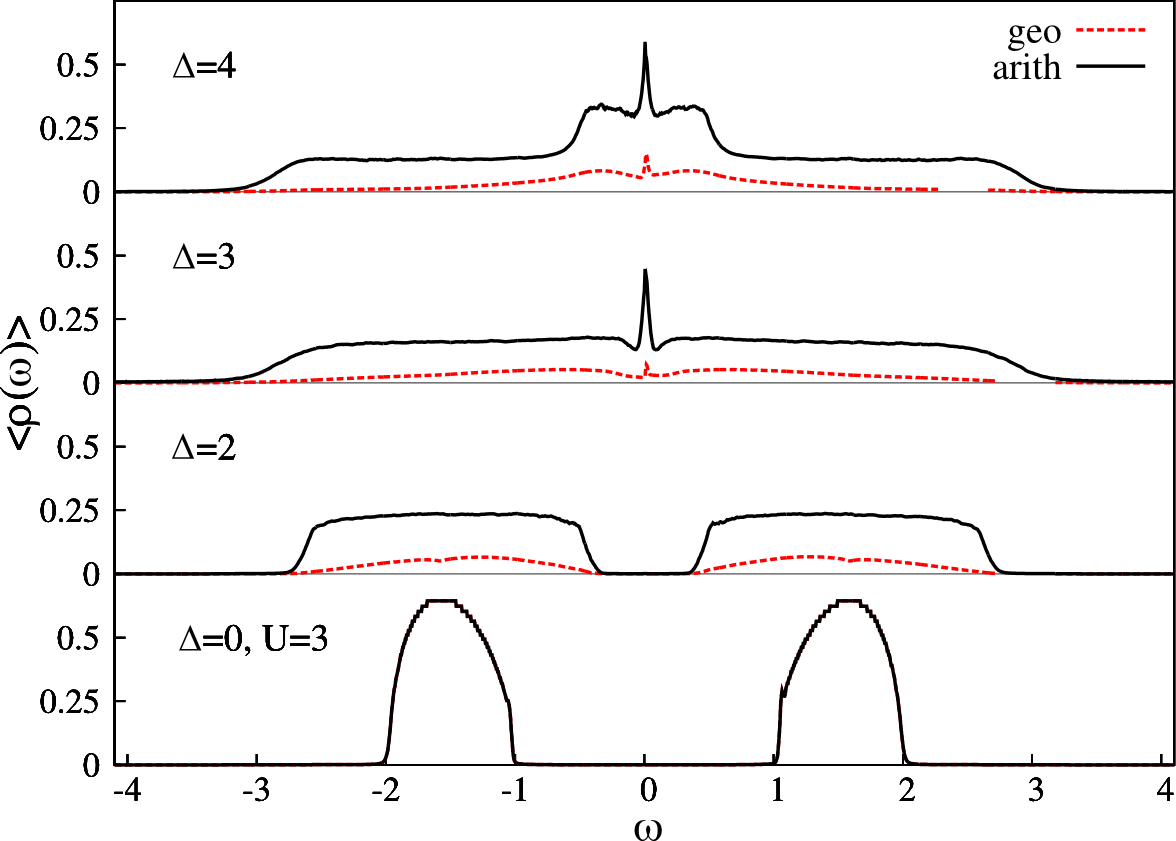}  
\caption{(Color online) Evolution of the arithmetically and geometrically averaged spectral function with increasing disorder strength $\Delta$ ($\Delta=0,2,3,4$) for fixed interaction strength $U=3$. Parameters are $\eta=10^{-3}$, $\mu=U/2$, $W_0=1$, and $K=3$.}
\label{fig_box_delocal_fixU}
\end{figure}

We will discuss how the homogeneous system in the Mott insulating state is affected by disorder. In Fig.~\ref{fig_box_delocal_fixU} the evolution of the geometrically and arithmetically averaged spectral function is presented at fixed interaction strength $U=3$. 
In the homogeneous system the corresponding state is Mott insulating. The Mott excitation gap persists for small disorder strengths. From the spectrum at $\Delta=2$ we clearly see that the Mott gap is reduced by disorder. This can be understood by a redistribution of states into the gap. At $\Delta=3$ the gap is completely filled. The geometric average is finite in the core of the band, but zero in the outer parts of the spectrum. This suggests that a metallic phase is obtained. This is confirmed by a proper investigation (which is not shown here) of the evolution of the PDF of the LDOS in the limit $\eta\rightarrow 0$ to rule out localized states at the Fermi level. Hence, the increase of the disorder strength causes a delocalization of the local  magnetic moments of the Mott insulator and the system undergoes a transition from the Mott insulator to a disordered metal. This result is in agreement with findings obtained by TMT.\cite{Byczuk2005} Moreover, we note that the Kondo resonance corresponding to coherent quasi-particle excitations is present in the disordered metal. 

\begin{figure}[t]
\centering
\includegraphics[width=0.48\textwidth]{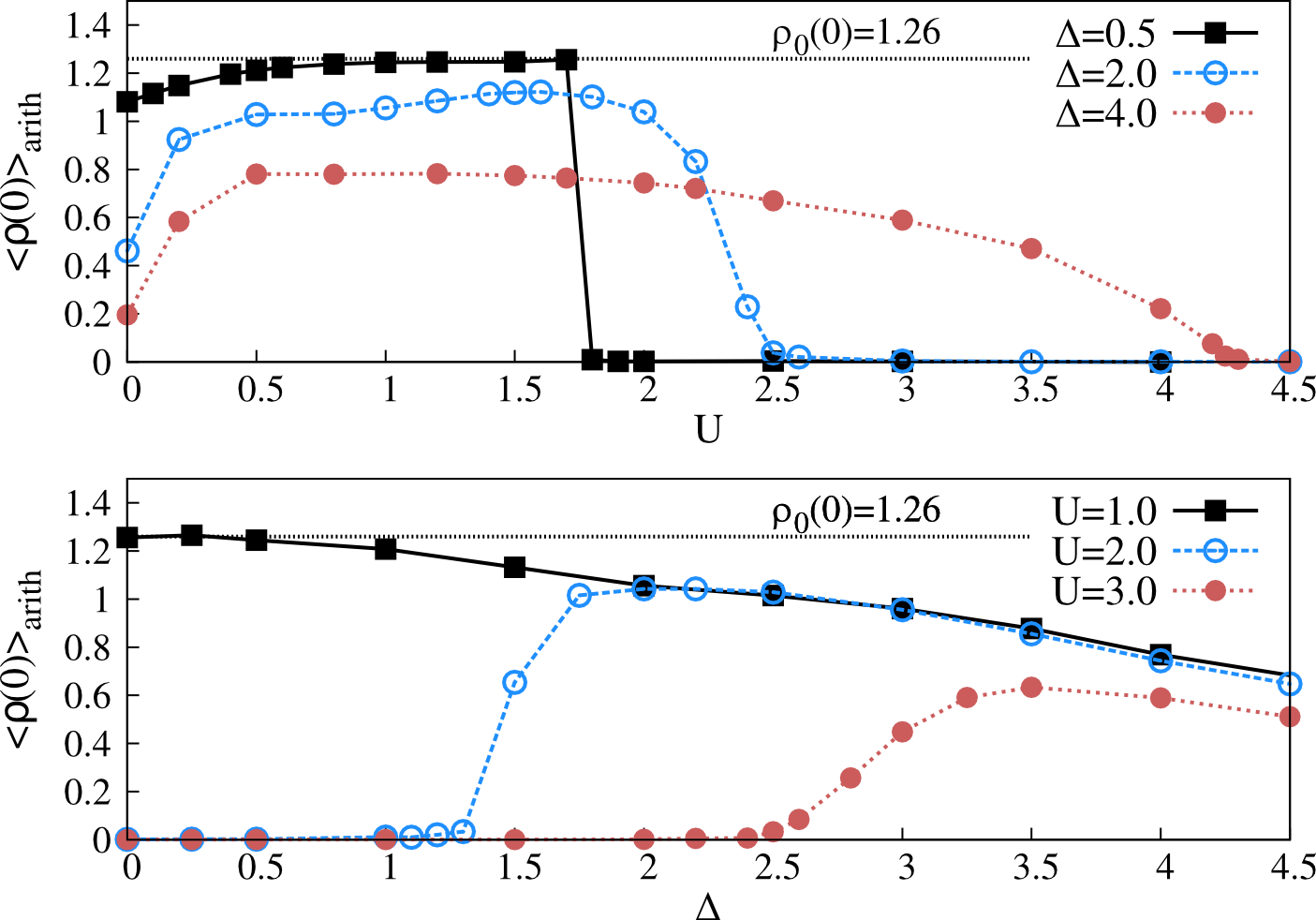}  
\caption{(Color online) Arithmetically averaged LDOS at the Fermi level $\langle\rho(0)\rangle_{\mbox{\tiny arith}}$ for three different values of the disorder strength $\Delta$: $\Delta=0.5$ (solid squares), $\Delta=2$ (open circles) and  $\Delta=3$ (solid circles) in the upper panel. In the lower panel the arithmetically averaged LDOS is plotted as a function of the disorder strength for three different values of the interaction strength: $U=1$ (solid squares), $U=2$ (open circles) and  $U=3$ (solid circles) in the upper panel. Parameters are $\mu=U/2$,  $\eta=10^{-3}$, $W_0=1$, and $K=3$.}
\label{fig_box_metallicity}
\end{figure}

The question how the correlation-induced metal-insulator transition is affected by randomness arises naturally. In previous TMT studies\cite{Byczuk2005} it was observed that the Luttinger theorem, i.e. the pinning of the LDOS at its non-interacting value\cite{MullerHartmann1989}, is not fulfilled in the presence of disorder. Moreover, it was shown that the metallicity, given by the DOS at the Fermi level, grows with increasing interaction strength so that the Luttinger theorem is nearly fulfilled again.\cite{Byczuk2005} Sufficiently strong interactions hinder a decay of the quasi-particle excitations. Further increase of the interaction strength leads to a rather abrupt transition to the Mott insulating phase. Within statistical DMFT we observe a similar behavior as given in the upper panel of Fig.~\ref{fig_box_metallicity}, where the arithmetic average of the LDOS at the Fermi level is plotted as a function of the interaction for three different disorder strengths.
For $U=0$ the arithmetic average LDOS decreases for higher disorder strengths as expected. As a consequence of increasing interaction we observe an increase of the metallicity for each value of the considered disorder strengths in agreement with the TMT results. For $\Delta=0.5$ the metallicity increases gradually until it suddenly drops to zero for $U_c=1.8$. This behavior indicates a first order transition in agreement with the observed hysteresis region within TMT.\cite{Byczuk2005} However, no sudden transition to the Mott insulator occurs for higher disorder strengths as for example displayed for $\Delta=2$ and $\Delta=4$ but a rather smooth decline. Our data does not allow for answering the question whether the decline corresponds to a second order transition or a crossover, whereas the previous TMT analysis suggests a crossover.\cite{Byczuk2005} 

The increase of metallicity is conform with the observed influence of the interaction strength on the PDF of the LDOS as shown in Fig.~\ref{fig_pdf_vary_U} for $\Delta=4$. The two regimes of distinct behaviors of the PDF upon increase of the interaction  discussed previously are not reflected in the metallicity. The metallic phase obviously features non-trivial properties, which might be associated to the non-Fermi liquid phase representing the precursor of the Anderson-Mott transition.\cite{Dobrosavljevic1997} Future research is needed to substantiate this possibility and the additional determination of the PDF of further local observables such as the quasi-particle decay rate might be necessary to properly characterize this phase. 

The lower panel of Fig.~\ref{fig_box_metallicity} displays the metallicity as a function of disorder strength for three different values of the interaction. The finite metallicity of the homogeneous system found for small interaction remains for increasing disorder strength, but reduces gradually. Starting within the Mott insulating phase of the homogeneous system, an increase of the interaction strength results in a finite metallicity from some critical value on, which confirms the disorder-induced delocalization.     

\begin{figure}[t]
\centering
\includegraphics[width=0.48\textwidth]{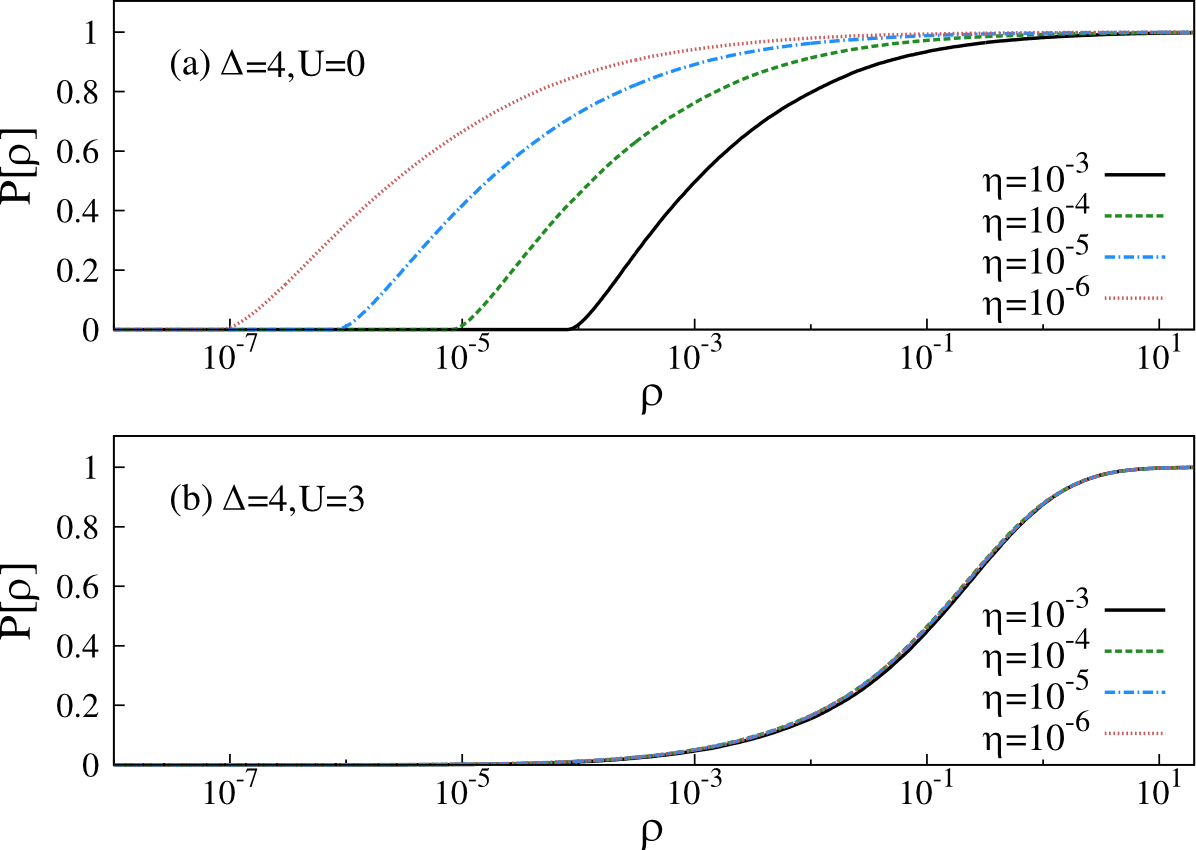}  
\caption{(Color online) Evolution of the cumulative PDF at the Fermi level $P[\rho(0)]$ with decreasing $\eta$ for the (a) non-interacting system and (b) in presence of interactions, $U=3$, at fixed disorder strength $\Delta=4$. Parameters are $\mu=U/2$, $W_0=1$, and $K=3$.}
\label{fig_deloc_by_U}
\end{figure}

Next we discuss how Anderson localized states are influenced by interactions. Panel (a) of Fig.~\ref{fig_deloc_by_U} shows the evolution of the PDF of the LDOS at the Fermi level upon $\eta\rightarrow 0$ for the non-interacting system with $\Delta=4$ and $K=3$. This state is Anderson insulating. In panel (b) the corresponding evolution is shown for finite interactions, namely $U=3$. Clearly, the PDF saturates corresponding to an extended state. The interaction obviously causes a transition from the Anderson insulator to the disordered metal. Thus, a second delocalization transition caused by the increase of the interaction strength is identified, which is in agreement with the corresponding TMT analysis.\cite{Byczuk2005} Now we are also able to understand the dramatic effect of the introduced correlations on the PDF of the LDOS as displayed in panel (a) of Fig.~\ref{fig_pdf_vary_U}, since the system is in the metallic state already for $U=0.5$.
  
\begin{figure}[t]
\centering
\includegraphics[width=0.48\textwidth]{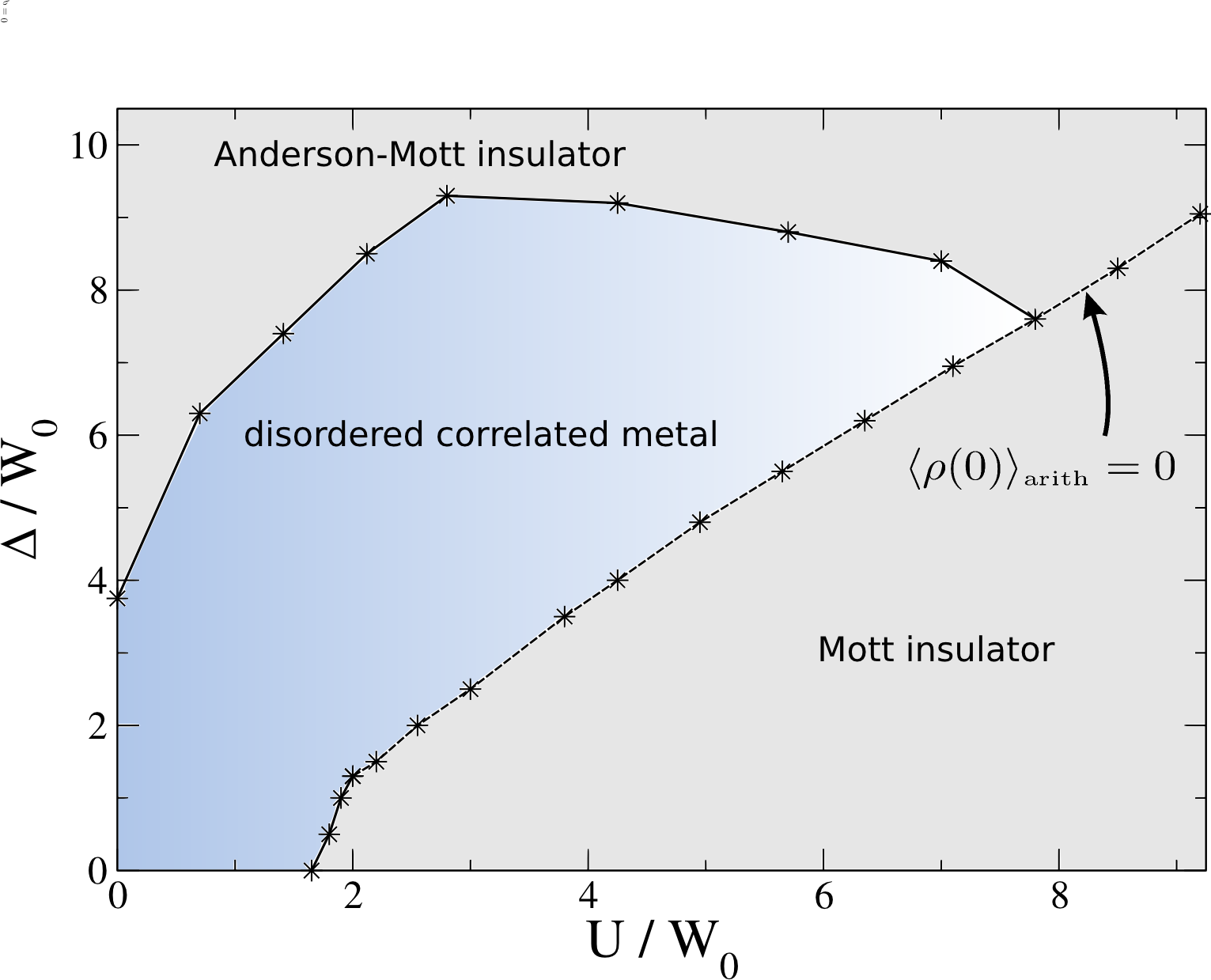}  
\caption{(Color online) Paramagnetic ground state phase diagram of the half-filled Anderson-Hubbard model calculated by means of statistical DMFT. It consists of Anderson-Mott insulator, Mott insulator and paramagnetic disordered metal. Parameters are $\mu=U/2$, $W_0=1.0$, and $K=3$.}
\label{fig_pd_box}
\end{figure}
Our systematic analysis results in the paramagnetic phase diagram depicted in Fig.~\ref{fig_pd_box}. Both above described delocalization processes are clearly visible, as can be judged from the extent of the metallic phase. For strong interactions a Mott insulator occurs, which is continuously connected to the Anderson-Mott insulator for strong disorder. That means, the  Mott insulating state can be continuously tuned into an Anderson-Mott insulating state by changing the disorder and the interaction strength. The dashed line corresponds to a vanishing arithmetic average of the LDOS at the Fermi level, as a reasonable distinction between the Mott insulator and the Anderson-Mott insulator. Within TMT studies a distinction was achieved by the criterion of vanishing Hubbard subbands\cite{Byczuk2005} or whether the quasi-particle weight drops to zero for all on-site energies or not.\cite{Aguiar2009} Here, we see that the Anderson-Mott insulator and the Mott insulator are separated both by the metallic phase and for strong interactions by a crossover line $\Delta\approx U$. Future research is needed to quantify the critical behavior of the transition from metall to Mott insulator for higher disorder  
strengths. Statistical DMFT investigations in combination with more powerful and accurate impurity solvers such as the numerical renormalization group\cite{Krishna-Murthy1980,Hofstetter2000,Bulla2008} could give valuable insight in this respect.

It is interesting to compare our statistical DMFT phase diagram with the phase diagram obtained by means of the TMT.\cite{Byczuk2005,Aguiar2009} The overall structure is reproduced. Within both methods one finds a metallic core for intermediate strengths of both the disorder and the interaction. Also the shape of the metallic core is in agreement. The two delocalization processes are obtained within both methods. However, on a quantitative level the methods differ substantially. In the non-interacting system, the TMT predicts the Anderson localization to occur at a critical disorder strength $\Delta_c \approx 1.8$.\cite{Byczuk2005} Within statistical DMFT the transition is found for significantly higher disorder strength. For the connectivity $K=3$ the critical disorder strength $\Delta_c \approx 3.75$ results in a discrepancy of a factor $\sim2$. We remark that within TMT no fluctuations due to disorder are incorporated. Furthermore, the choice of the mean, i.e. the geometric average, is to some degree arbitrary and has a strong influence on the extent of the metallic phase.\cite{Souza2007} This discrepancy was already discussed by Alvermann and Fehske for the non-interacting system.\cite{Alvermann2005} Therein it was also shown that TMT misses the re-entrant behavior of the mobility edge that was found by means of the local distribution method. Moreover, our discussion above shows that within statistical DMFT the extent of the metallic phase depends on the connectivity $K$ as it is expected from a physical reasoning. For higher connectivities, within  statistical DMFT, this discrepancy to TMT findings will even be consolidated, since the critical disorder strength for Anderson or Anderson-Mott localization will be higher. As a consequence, the crossover line between the metallic phase and the Mott insulating phase exists for much larger disorder strengths than found within TMT. 

On the other hand, it should be mentioned that within the TMT investigation the numerical renormalization group was employed as impurity solver. Therewith, the Mott transition is determined essentially numerically exactly. In contrast, within the here presented statistical DMFT approach the approximate MPT was used, which is known to differ from the most accurate values known.\cite{Bulla2001}

To sum up the comparison, our findings indicate that criticism regarding the detection of localization within TMT\cite{Alvermann2005,Song2007,Souza2007} is justified on a quantitative level. However, our investigation supports that statistical DMFT and TMT lead to qualitatively agreeing results in general. In this sense, the TMT represents a valuable investigation tool for strongly correlated, disordered systems. An important advantage of the TMT is the considerable smaller computational effort, which enables the use of more powerful impurity solvers than it is feasible to date within statistical DMFT.

\section{Conclusion}\label{summary}

The statistical DMFT was applied to the box disordered Anderson-Hubbard Hamiltonian. We have confirmed that statistical DMFT is able to describe localization transitions in the presence of interactions and disorder. Localization is judged by the evolution of the PDF of the LDOS in the limit $\eta \rightarrow 0$, which enables the investigation of the Anderson-Mott transition. 

It was found that the presence of interactions in the Anderson insulating phase causes delocalization for intermediate disorder strengths. On the other hand, also disorder delocalizes the Mott insulating phase by the redistribution of states into the Mott gap. However, if the disorder or respectively the interaction is strong enough, no metallic phase occurs anymore. Consequently, the Mott insulator and the Anderson-Mott insulator are continuously connected. The resulting paramagnetic ground state phase diagram was found to be in qualitative agreement with TMT results. However, considerable quantitative differences between statistical DMFT and TMT were found regarding the extent of the metallic phase.

Our analysis of the probability distribution functions of the local density of states in the interacting case revealed a non monotonic dependence on the interaction strength. The most striking feature is an emerging second peak in the probability distribution function which was not observed before. In particular, the log-normal distribution does not serve as a suitable approximation in the interacting case. 

Our results may be not only relevant for the solid state community but also for future experiments with ultracold fermions in disordered optical lattices,\cite{Aspect2009, Sanchez-Palencia2010} in which the disorder strength and the interaction can be controlled and tuned independently.    

\section*{Acknowledgments}

We acknowledge useful discussions with U.~Bissbort and J.~Wernsdorfer. This work was supported by the Deutsche Forschungsgemeinschaft (DFG) via Forschergruppe FOR 801. Computations were performed at the Center for Scientific Computing (CSC) at the Goethe University Frankfurt. K.B. acknowledges support by the grant TRR80 of the DFG and the grant No. N~N202~103138 by the Polish Ministry of Science and Education.


\begin{thebibliography}{99}
\bibitem{comment1} Even on a real Bravais lattice restricted to the leading contribution, the first equation would extend over $z$ neighbors, but all subsequent equations would incorporate $z-1$ or less summands.
\end{thebibliography}
\end{document}